\documentclass{elsart3}

\usepackage{graphicx}

\begin{document}

\begin{frontmatter}



\title{Universal magnetic excitation spectrum in cuprates}


\author[BNL]{J. M. Tranquada,}
\author[BNL,ISIS]{H. Woo,}
\author[ISIS]{T. G. Perring,}
\author[Tohoku]{H. Goka,}
\author[BNL]{G. D. Gu,}
\author[BNL]{G. Xu,}
\author[Tohoku]{M. Fujita,}
\author[Tohoku]{and K. Yamada}

\address[BNL]{Physics Department, Brookhaven National Laboratory, Upton,
NY 11973-5000, USA}
\address[ISIS]{ISIS Facility, Rutherford Appleton Laboratory, Chilton,
Didcot, Oxon OX11 0QX, UK}
\address[Tohoku]{Institute for Materials Research, Tohoku University,
Sendai 980-8577, Japan}

\begin{abstract}
We have recently used inelastic neutron scattering to measure the
magnetic excitation spectrum of La$_{1.875}$Ba$_{0.125}$CuO$_4$ up to 200
meV.  This particular cuprate is of interest because it exhibits static
charge and spin stripe order.  The observed spectrum is remarkably
similar to that found in superconducting YBa$_2$Cu$_3$O$_{6+x}$ and
La$_{2-x}$Sr$_x$CuO$_4$; the main differences are associated with the
spin gap.  We suggest that essentially all observed features of the
magnetic scattering from cuprate superconductors can be described by a
universal magnetic excitation spectrum multiplied by a spin gap function
with a material-dependent spin-gap energy.
\end{abstract}

\end{frontmatter}

\section{Introduction}
\label{}

Given the prominent role of antiferromagnetism in the typical phase
diagram for cuprate superconductors, it is commonly believed that
antiferromagnetic spin fluctuations play a significant role in the
mechanism of superconductivity.  Experimentally, however, it has been
difficult to establish a universal trend for the magnetic excitations
that applies across all hole-doped cuprate families.

One fairly broad experimental trend involves the appearance of a magnetic
``resonance'' peak at roughly 40 meV in the superconducting
state.  Centered at the antiferromagnetic wave vector ${\bf Q}_{\rm AF}$,
this feature has been observed in YBa$_2$Cu$_3$O$_{6+x}$
\cite{bour98,dai01}, Bi$_2$Sr$_2$CaCu$_2$O$_{8+\delta}$ \cite{fong99}, and
Tl$_2$Ba$_2$CuO$_6$ \cite{fong99,he02}.  A problem with this trend is that
an analogous feature has not been observed in the family of cuprates
associated with La$_{2-x}$Sr$_x$CuO$_4$.

Our recent neutron scattering results for La$_{1.875}$Ba$_{0.125}$CuO$_4$
(LBCO) \cite{tran04} allow one, for the first time, to identify a
universal magnetic excitation spectrum for the cuprates.  As we discuss
below, the observed excitations show dispersions quite similar to those
found in several recent studies of YBa$_2$Cu$_3$O$_{6+x}$ (YBCO)
\cite{hayd04,rezn03,stoc04b,pail04}.  The major difference is with
respect to the temperature and frequency dependence associated with the
spin gap.  We note that the temperature-dependent effects in
La$_{2-x}$Sr$_x$CuO$_4$ (LSCO) are associated with a small spin gap, and
thus occur at an energy scale where the magnetic excitations are
incommensurate \cite{maso96,lake99,tran04b,gila04,chri04}.  We suggest
that all of the observed behavior can be described by a single
phenomenological model in which there is a universal magnetic excitation
spectrum with the intensity multiplied by a gap function, with the spin
gap energy roughly correlated with the superconducting transition
temperature, $T_c$, and varying among cuprate families.

Another significant feature of our LBCO sample is that it exhibits static
charge and spin stripe order \cite{fuji04}.   Cartoons of the two
possible stripe domains are shown in Fig.~1. The stripe order competes
with superconductivity \cite{ichi00}, so that the
$T_c$ of our sample is less than 6~K, well below the measurement
temperature of 12~K.  As our measurements are in the normal state, they
indicate that the universal features of the excitation spectrum cannot
depend on the existence of a coherent superconducting state.  We discuss
interpretations of the spectrum in terms of the quantum excitations of
finite spin clusters such as 2-leg antiferromagnetic ladders.  We also
comment on alternative explanations in terms of fermiology.

\begin{figure}[t]
\leftline{\includegraphics[width=3in]{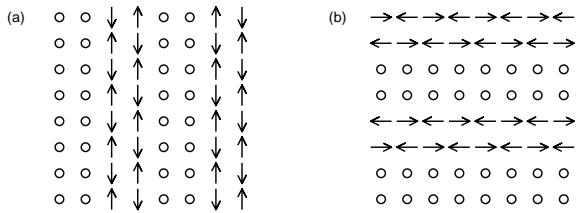}}
\medskip
\caption{Cartoons of charge and spin stripes, showing two types of
stripe domain.  Arrows indicate Cu spins; circles indicated hole-rich
regions.}
\label{fg:stripes} 
\end{figure}

\section{Neutron scattering experiment on La$_{1.875}$Ba$_{0.125}$CuO$_4$}

Neutron scattering measurements have established that stripe order
appears in LBCO below 50 K
\cite{fuji04}.  For the present experiment, four large crystals (each 8
mm $\phi\ \times$ 50 mm) with a total mass of 58 g were grown at
Brookhaven.  After coaligning the crystals at the JRR-3M reactor in
Tokai, Japan, the sample was transported to the ISIS spallation source. 
The experiment was performed on the MAPS spectrometer, a direct-geometry
time-of-flight spectrometer with a large position-sensitive area
detector.  The sample was aligned with the $c$ axis
parallel to the incident beam direction.  Measurements were performed
with incident energies of 80, 240, and 500 meV.  

The results have been reported in Ref.~\cite{tran04}.  At low energies
($\le10$ meV), we observe magnetic excitations peaked at the the
wave vectors of the magnetic superlattice peaks \cite{fuji04}.  These are
essentially the same incommensurate wave vectors at which the low-energy
magnetic excitations are found in La$_{2-x}$Sr$_x$CuO$_4$ \cite{cheo91}. 
With increasing energy, the excitations disperse inwards towards 
${\bf Q}_{\rm AF}$, with no obvious outwardly dispersing excitations;
similar results have recently been reported for optimally-doped
La$_{2-x}$Sr$_x$CuO$_4$ \cite{chri04}.  These excitations merge at 
${\bf Q}_{\rm AF}$ when the energy reaches $\sim50$~meV.  At higher
energies, the scattering disperses outwards, away from 
${\bf Q}_{\rm AF}$.  At a given energy, the shape of the scattering in
the $(h,k,0)$ zone of reciprocal space forms a square with its corners
rotated by 45$^\circ$ relative to the square formed by the low-energy
incommensurate wave vectors.  Through their {\bf Q} dependence, we have
been able to identify magnetic excitations up to $\sim200$~meV; the
scattering strength appears to fall off rapidly above that energy.

\section{Similarity to YBa$_2$Cu$_3$O$_{6+x}$}

The dispersion of the magnetic excitations described above is extremely
similar to new observations on YBa$_2$Cu$_3$O$_{6.6}$ (YBCO6.6) by Hayden
{\it et al.} \cite{hayd04} obtained on the same spectrometer.  If one
multiplies the energy scale for the features in LBCO by
$\sim\frac23$ (the approximate ratio of the ``resonance'' energies) then
they match up quite well with those reported for YBCO6.6.  Of course,
there are certain obvious differences.  The YBCO6.6 sample was measured
in the superconducting state, whereas LBCO was studied in the normal
state.  The LBCO has static magnetic order, while the YBCO6.6 has a
spin gap of $\sim20$ meV.  Despite these differences, the similarities
suggest that there may be a universal magnetic spectrum common to the
hole doped cuprates.  

A study of partially detwinned YBCO6.5 by another group, Stock {\it et
al.} \cite{stoc04b}, confirms the nature of the dispersion.  They report
that the scattering at high energies forms more of a circle than a square
in 2D reciprocal space; however, such discrepancies are of less
significance than the universal features that are apparent.  A similar
dispersion with energy has also been seen in studies of YBCO closer to
optimal doping \cite{rezn03,pail04}.

\section{Weakly-coupled ladder model}

Given that our LBCO sample exhibits stripe order \cite{fuji04}, it is
natural to look for an explanation of the universal magnetic spectrum in
terms of the inhomogeneous antiferromagnetic correlations associated with
that order \cite{zaan01,kive03,sach91}.  Several years ago, Batista,
Ortiz, and Balatsky \cite{bati01} proposed that 
the magnetic resonance observed in YBCO might be
understood in terms of the excitations of an incommensurate
spin-density-wave state.  They made an analogy with the excitation
spectrum observed in (diagonally) stripe-ordered
La$_{1.67}$Sr$_{0.33}$NiO$_4$ \cite{boot03,bour03}.  The latter spectrum
has since been shown to be well described by linear spin-wave theory
\cite{boot03,kane01,krug03,carl04}.

The problem with this picture is that the
observed dispersions are different from the predictions of linear spin
wave theory applied to ordered stripes, as has been pointed out by
Bourges {\it et al.} \cite{bour00} in their neutron scattering study of
YBCO6.85.  Spin waves disperse isotropically from the
incommensurate modulation wave vectors, with similar intensities for spin
waves dispersing towards and away from the antiferromagnetic wave
vector ${\bf Q}_{\rm AF}$.  In contrast, there is no sign of any outward
dispersing excitations in YBa$_2$Cu$_3$O$_{6.85}$, and only dispersion
towards ${\bf Q}_{\rm AF}$ is seen.  Our results for LBCO confirm that the
excitations of a stripe-ordered cuprate differ from the predictions of
spin-wave theory \cite{tran04}.

Looking at Fig.~1, one can see that, at least in this cartoon version of
stripe order, the hole-rich stripes separate 2-leg antiferromagnetic spin
ladders.  An isolated ladder with superexchange coupling $J$ between
nearest-neighbor spins has an excitation gap of $J/2$ \cite{dago96}.  In
the ground state the spins tend to form singlet pairs, and a triplet
excitation can propagate along the length of a ladder \cite{barn94}. 
We found that the dispersion of an isolated 2-leg ladder with
$J=100$~meV gives a good description of our measurements above 50 meV. 
Such a model suggests that the gap at the antiferromagnetic wave vector
is associated with singlet spin correlations.

To simultaneously describe the incommensurate excitations at lower
energies, one must take into account the coupling between the ladders. 
Calculations of the spectrum under the assumption of weak coupling
between the ladders have been reported by Vojta and Ulbricht
\cite{vojt04} and by Uhrig, Schmidt, and Gr\"uninger \cite{uhri04}. 
The former paper was the first to show that the calculated spectrum for
weakly-coupled ladders gives a good description of the dispersion in
LBCO, while the latter paper included cyclic ring exchange in the
ladder Hamiltonian and showed that the model can give quantitative
agreement with the absolute scale of the observed magnetic scattering. 
Some features of the dispersion were anticipated by an earlier
calculation by Dalosto and Riera \cite{dalo00}.  Good agreement with the
measurements is also obtained in a calculation of fluctuations about a
mean-field stripe-ordered state by Seibold and Lorenzana \cite{seib04}.

It seems worthwhile to note that there are experimental precendents for
the coexistence of quantum excitations with static magnetic order.  Such
behavior has been seen in compounds containing $S=1$ spin chains weakly
coupled together through magnetic rare-earth ions \cite{yoko98}, and in
$S=1$ spin chains doped with holes \cite{xu00b}.

\section{Fermiology}

An alternative approach to the origin of the
magnetic excitations in the superconducting cuprates is based on Fermi
liquid theory.  One calculates the spin susceptiblity based on the ability
of conduction electrons to scatter across the Fermi surface from filled to
empty states \cite{si93,norm00}.  From this perspective, the resonance
feature is commonly tied to the shape of the electronic Fermi surface and
the $d$-wave nature of the superconducting gap \cite{lu92,bulu96}.  Such
calculations are capable of describing (or fitting) features
of the observed magnetic excitations, such as
incommensurability \cite{brin99,kao00,yama01,onuf02,ito02,schn04}.

There are various features of this approach that we find to be
problematic.  There are significant doubts concerning the
appropriateness of such calculations in the normal state, especially
in the underdoped regime.  It is also unclear how to explain
charge-stripe order in this picture.  (For more discussion of such
issues, see Kivelson {\it et al.} \cite{kive03}.)  As our LBCO sample is
in the normal state and has charge stripe order, calculations based on
fermiology do not seem to be relevant.  Given the similarity between the
magnetic dispersion in our LBCO sample and that found in YBCO, we
question the relevance of fermiology for obtaining a global understanding
of the magnetic response in the cuprates.

\section{Universal magnetic spectrum and the superconducting spin gap}

The LBCO results indicate that the occurrence of commensurate inelastic
scattering does not require superconductivity.  Thus, what seems to be
special about the ``resonance'' phenomenon is its temperature dependence,
and not that the excitation is commensurate.  
In the superconducting state, the enhanced magnetic signal always appears
just above the spin gap.  Thus, it appears that one might be able to
describe all of the neutron scattering results obtained so far on a
variety of hole-doped cuprates with a universal spectral function
multiplied by a suitable gap function.  The energy scale for the spectrum
and the size of the gap are clearly material and doping dependent. 
Experimentally, the size of the spin gap is roughly correlated with
$T_c$, while the gap at the commensurate wave vector may be an upper
limit to the spin gap.  Such behavior is consistent with a
superconducting mechanism in which spin pairing drives hole pairing
\cite{arri04}.

Motivated by mean-field calculations \cite{koni04}, we have calculated a
model spin  susceptibility based on weakly coupled 2-leg ladders using
\begin{equation}
  \chi = \chi_{\rm ladd}/[1+J_\perp\sin^2(4\pi q_\perp)\chi_{\rm ladd}],
\end{equation}
where we take
\begin{equation}
  \chi_{\rm ladd} = {1\over\hbar\omega(q_\|)}
   \cos^2(\pi q_\|)
   \cos^2(\pi q_\perp) F(\omega),
\end{equation}
with $\omega(q_\|)$ from \cite{barn94}, and
\begin{equation}
  F(\omega) = {1\over E-\hbar\omega(q_\|)+i\gamma} -
              {1\over E+\hbar\omega(q_\|)+i\gamma},
\end{equation}
with $q_\| = h - {\textstyle\frac12}$, $q_\perp= k-{\textstyle\frac12}$,
and $E = \hbar\omega$. Here, $J_\perp$ is the effective coupling between
ladders, which we set to $0.13J$, where $J$ is the superexchange within a
ladder.  Figure 2(b) shows the spectrum of $\chi''$ calculated along the
path shown in Fig.~2(a), averaged over the two stripe orientations shown
in Fig.~1.  We take this to be a rough model of the excitation spectrum
that we have measured in LBCO.  

\begin{figure*}[t]
\centerline{\includegraphics[width=6in]{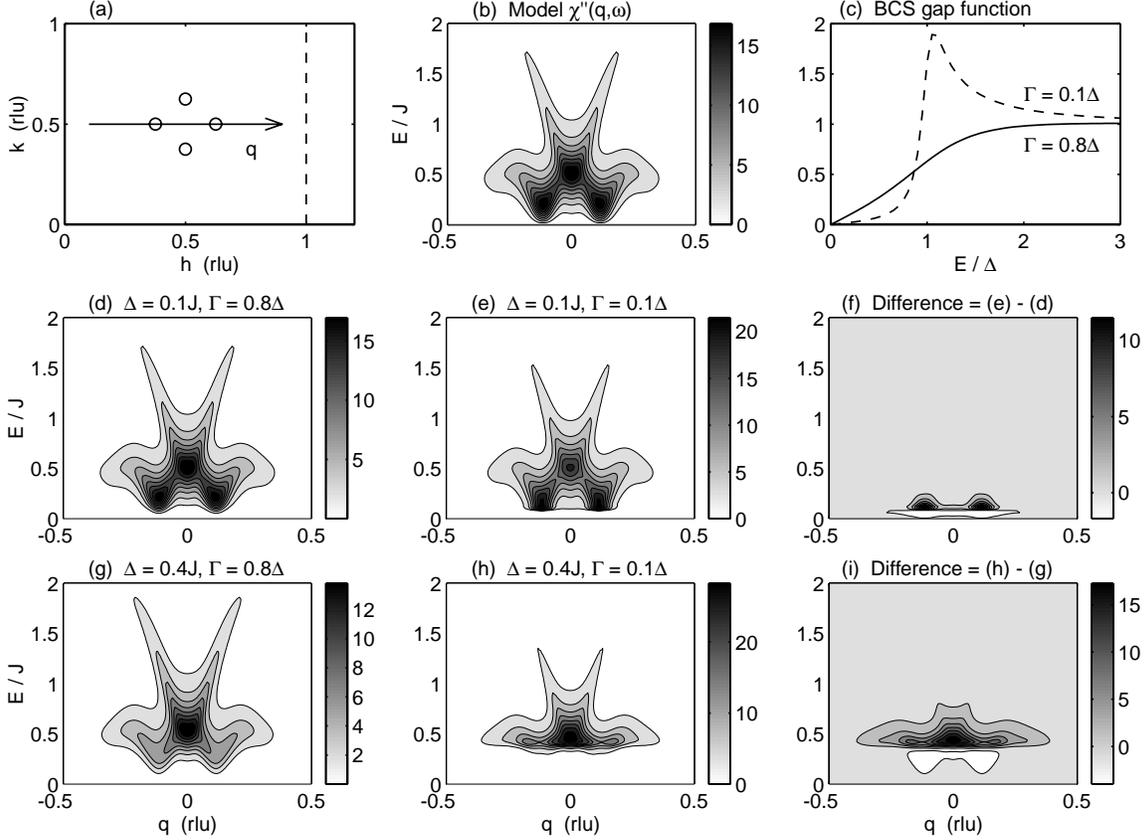}}
\medskip
\caption{(a) Arrow indicates path in reciprocal space used in following
panels.  (b) Model of $\chi''$ as discussed in text.  (c) BCS gap
function for two choices of $\Gamma$.  (d) Product of model $\chi''$ and
gap function with $\Delta=0.1(1+i0.8)J$, and (e) $\Delta=0.1(1+i0.1)J$. 
(f) Difference of (e) and (d).  (g) Similar to (d), with
$\Delta=0.4(1+i0.8)J$, and (h) $\Delta=0.4(1+i0.1)J$.  (i) Difference of
(h) and (g).}
\label{fg:model} 
\end{figure*}

The major difference between the experimental spectrum for LBCO and that
for superconducting cuprates is the spin gap observed below $T_c$.  To
model the spin gap, we multiply $\chi''$ by a gap function.  We
arbitrarily choose to use the BCS gap function,
${\rm Re}(E/\sqrt{E^2-\Delta^2})$, with $\Delta = \Delta_0+i\Gamma$.  The
gap function alone is plotted in Fig.~2(c): for $\Gamma=0.1\Delta_0$ the
depression of weight below the gap and the pile up of weight above the gap
are apparent, while for $\Gamma=0.8\Delta_0$ the gap is overdamped.  We
suggest that the overdamped gap mimics the normal-state response of the
doped cuprates.

To mimic optimally and over doped LSCO, we choose $\Delta_0 = 0.1J$,
considerably smaller than the ladder gap energy of $0.5J$.  The model
calculations with overdamped and underdamped gaps are shown in Fig.~2(d)
and (e), respectively.  Taking these to represent normal state and
superconducting responses, we plot the difference in Fig.~2(f), which
shows a pile up of weight at incommensurate wave vectors, above a loss of
weight at lower energies.  The difference, restricted to incommensurate
scattering, is similar to experiment \cite{maso96,tran04b,gila04}.  In
contrast, we pick $\Delta_0=0.4J$ to mimic YBCO near optimal doping,
Fig.~2(g) and (h).  The difference, shown in Fig.~2(i), shows a large
gain at the commensurate position, with a loss of weight at lower
energies, as observed experimentally.

While this comparison is only qualitative, we believe it gives a
useful description of experiment.   A clear spin gap is
only observed in the superconducting state, but an overdamped gap may
characterize the normal state, at least in the underdoped regime.

\section{Acknowledgements}

Research at Brookhaven
is  supported by the Department of Energy's (DOE) Office  of Science
under Contract No.\ DE-AC02-98CH10886.  K.Y. and M.F. are supported by the
Japanese Ministry of Education, Culture, Sports, Science and Technology.
This work was supported in part by the USÐJapan Cooperative Research
Program on Neutron Scattering. 



\bibliographystyle{elsart-num}
\bibliography{lno,theory}







\end{document}